\documentclass{spie}
\usepackage{graphicx}
\usepackage{amsfonts}
\usepackage{amsmath}
\usepackage{amssymb}
\usepackage{xspace}
\title{Improved Performance of TES Bolometers using Digital Feedback}

\author{
 Tijmen de Haan\supit{a},
 Graeme Smecher\supit{a}\supit{b},
 Matt Dobbs\supit{a}
 \skiplinehalf
 \supit{a}Department of Physics, McGill University, 3600 Rue University, Montreal, Quebec H3A 2T8, Canada \\
 \supit{b}Three-Speed Logic, Inc., Vancouver, BC V6A 2J8, Canada
}
\authorinfo{
Further author information: (Send correspondence to T. de Haan) \\
T. de Haan: E-mail: tijmen@physics.mcgill.ca, Telephone: +1 (514) 398 4838 \\
G. Smecher: E-mail: gsmecher@threespeedlogic.com \\
M. Dobbs: E-mail: matt.dobbs@mcgill.ca \\
}

\newcommand{\degree}{$^\circ$}
\newcommand{\loopgain}{\ensuremath{\mathcal{L}}}
\newcommand{\fll}{shunt feedback\xspace}

\begin{document} 
\maketitle 

\begin{abstract}
Voltage biased, frequency multiplexed TES bolometers have become a widespread tool in mm-wave astrophysics. However, parasitic impedance and dynamic range issues can limit stability, performance, and multiplexing factors. Here, we present novel methods of overcoming these challenges, achieved through digital feedback, implemented on a Field-Programmable Gate Array (FPGA). In the first method, known as Digital Active Nulling (DAN), the current sensor (e.g. SQUID) is nulled in a separate digital feedback loop for each bolometer frequency. This nulling removes the dynamic range limitation on the current sensor, increases its linearity, and reduces its effective input impedance. Additionally, DAN removes constraints on wiring lengths and maximum multiplexing frequency. DAN has been fully implemented and tested. Integration for current experiments, including the South Pole Telescope, will be discussed. We also present a digital mechanism for strongly increasing stability in the presence of large series impedances,
 known as Digitally Enhanced Voltage Bias (DEVB).
\end{abstract}


\section{Introduction} 
\label{sec:intro}

\begin{figure}
  \centering
  \includegraphics[width=0.9\linewidth]{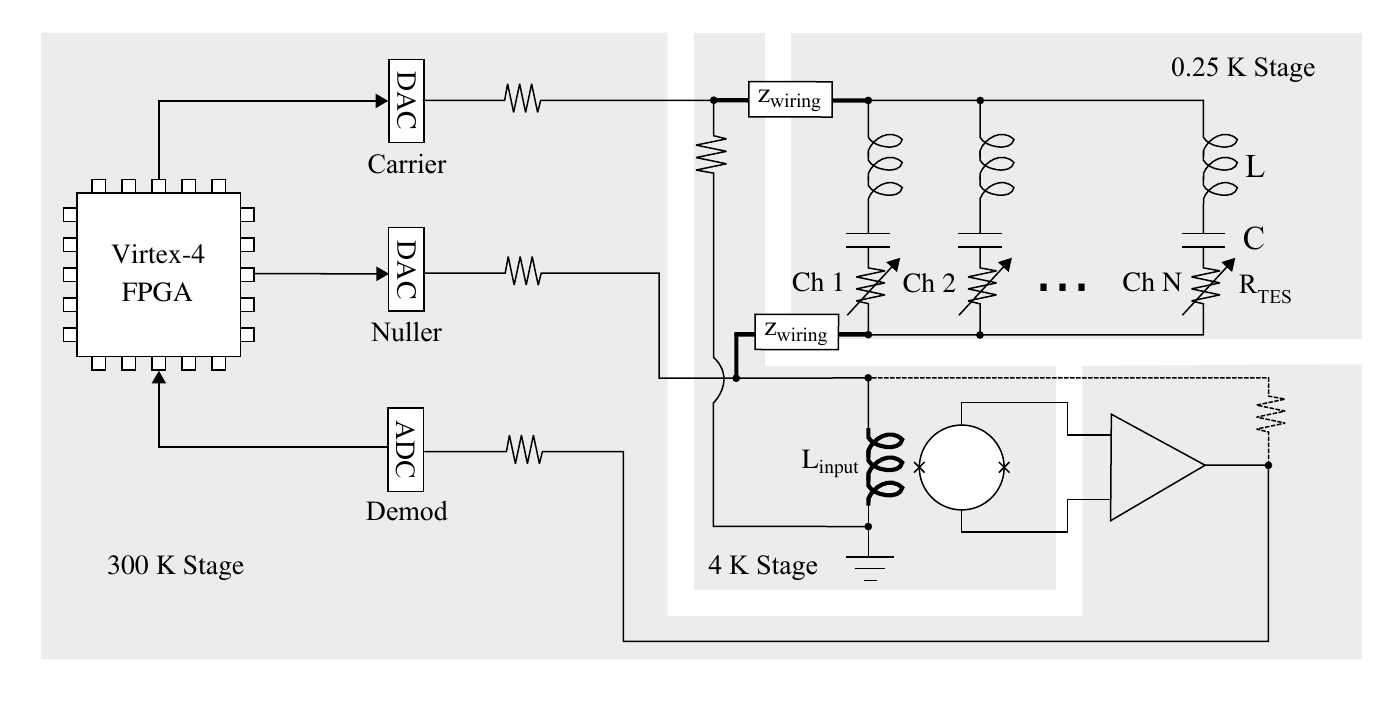}
  \caption{Simplified schematic showing the voltage biased, frequency multiplexed bolometer readout system. The digitized carrier waveform provides strong voltage bias to each TES at its respective resonant frequency, as set by the series LC filter. The SQUID provides a current measurement, which is amplified, digitized and demodulated. In previous operation, the digitized nuller signals provided static sinusoidal feedback, in addition to broadband feedback from the \fll. The latter, shown dashed, is omitted in DAN operation, where the nuller signal actively zeroes the SQUID current across the bolometer bandwidth. The thick lines highlight the main sources of series impedance to the bolometer.}
  \label{fig:readout}
\end{figure}

A large number of mm-wave astrophysics experiments \cite{dobbs11, schwan11, reichborn10, errard10} employ frequency multiplexed, voltage biased transition edge sensor (TES) bolometers to convert the incident radiation power into a current, which is measured by superconducting quantum interference devices (SQUIDs). A simplified schematic of this readout scheme is shown in Figure \ref{fig:readout}.

Due to the steep total power-resistance relation of the TES, a voltage bias results in strong electrothermal feedback (ETF) \cite{richards94}. In essence, the sum of optical and electrical power deposited on the bolometer is kept constant at the level dictated by the ETF loop gain. Series impedances to the TES spoil ETF and decrease stability, as we further explore in \S\ref{sec:measurements}. The main sources of series impedance are the input impedance of the SQUID and the inductance of the cryogenic wiring. 

In addition, the SQUID has a highly non-linear response and has limited dynamic range. Providing negative feedback to the SQUID linearizes it and suppresses its effective input impedance. This negative feedback was previously provided broadband using a SQUID flux locked loop shunt feedback\cite{dobbs11, dobbs12} (hereafter referred to as \fll). However, the \fll has several issues. First, it needs to provide strong negative feedback over the entire bandwidth, while maintaining stability. Managing phase shifts and rolling off the open loop gain to maintain the stability of this loop is a challenge \cite{lueker11}. This requirement currently limits the usable bandwidth for frequency multiplexing to $\sim 1.3~\mathrm{MHz}$ and restricts wiring lengths. In addition, the presence of a superconducting leg in the LCR comb (i.e. a latched TES) can cause the \fll to go unstable \cite{lueker11}.

Here we describe an alternative feedback scheme known as Digital Active Nulling (DAN), which does not suffer from these limitations, allowing a higher multiplexing factor, reduction of device parameter requirements, and improved stability. DAN is therefore a key technology for enabling multi-kilopixel bolometer arrays, for future ground-based, balloon-borne, and satellite applications.

DAN is similar to baseband feedback \cite{hartog11, hartog12, takei09} (BBFB), except for the electrical injection of the nulling signal, whereas BBFB is applied directly to the SQUID through a separate feedback coil. If the coupling constant (defined as the ratio of mutual inductance to the geometric mean of self inductances) of the input coil to the feedback coil were unity, this would be equivalent to DAN as far as the feedback loop is concerned. However, BBFB applications have so far had a small mutual inductance of the feedback coil to the input coil, in which case the increased linearity and dynamic range are preserved, but the SQUID input impedance suppression is equal to the coupling constant.

Furthermore, in \S\ref{sec:devb} we present a novel method of suppressing the effect of impedance in series with the TES known as Digitally Enhanced Voltage Bias (DEVB). Here, we once again take advantage of the small bandwidth of the TES and digitally correct for the voltage drop across the stray impedance; effectively providing strong voltage bias at each TES bias frequency separately.

Series impedances spoil the voltage bias. In the extreme limit where the series impedance dominates the total impedance, the TES is effectively current biased and is unstable to perturbations. An increase in current causes an increase in electrical power and hence in resistance, leading to an instability. As shown in Appendix \ref{app:stability}, for an effective series complex impedance $z_s$ and TES resistance $R_\mathrm{TES}$, the stability criterion is given by
\begin{equation}
  \label{eq:stability}
  \frac{d \log V_{\mathrm{TES}}}{d \log R_{\mathrm{TES}}} \leq \frac{|z_s|}{|z_s + R_\mathrm{TES}|} \, .
\end{equation}
Once the logarithmic slope of the transition exceeds the fractional effective series impedance, the TES will latch into a superconducting state, such that it is no longer useful as a detector. In \S \ref{sec:measurements}, we show that DEVB allows for stable operation in this regime. 

\section{Digital Active Nulling}
\label{sec:dan}

We have implemented DAN on the McGill digital frequency multiplexing (DfMux) platform \cite{dobbs08}. Each DfMux circuit board provides 8 digital-to-analog converters, 4 analog-to-digital converters, and a Xilinx Virtex-4 Field-Programmable Gate Array (FPGA) for signal processing tasks. Each digital-to-analog converter synthesizes a ``comb'' waveform at 25 MSPS, consisting of 16 sinusoidal terms with independent amplitude, phase, and frequency controls. The DfMux also provides 64 complex demodulator channels, and streams data from each of them over Ethernet at a programmable sampling rate which, for this work, we set to $191~\mathrm{Hz}$. The software stack consists of an embedded Linux system, running C code and controlled remotely with a user-friendly python API \cite{smecher12, story12}.

\begin{figure}
  \centering
  \includegraphics[width=0.9\linewidth]{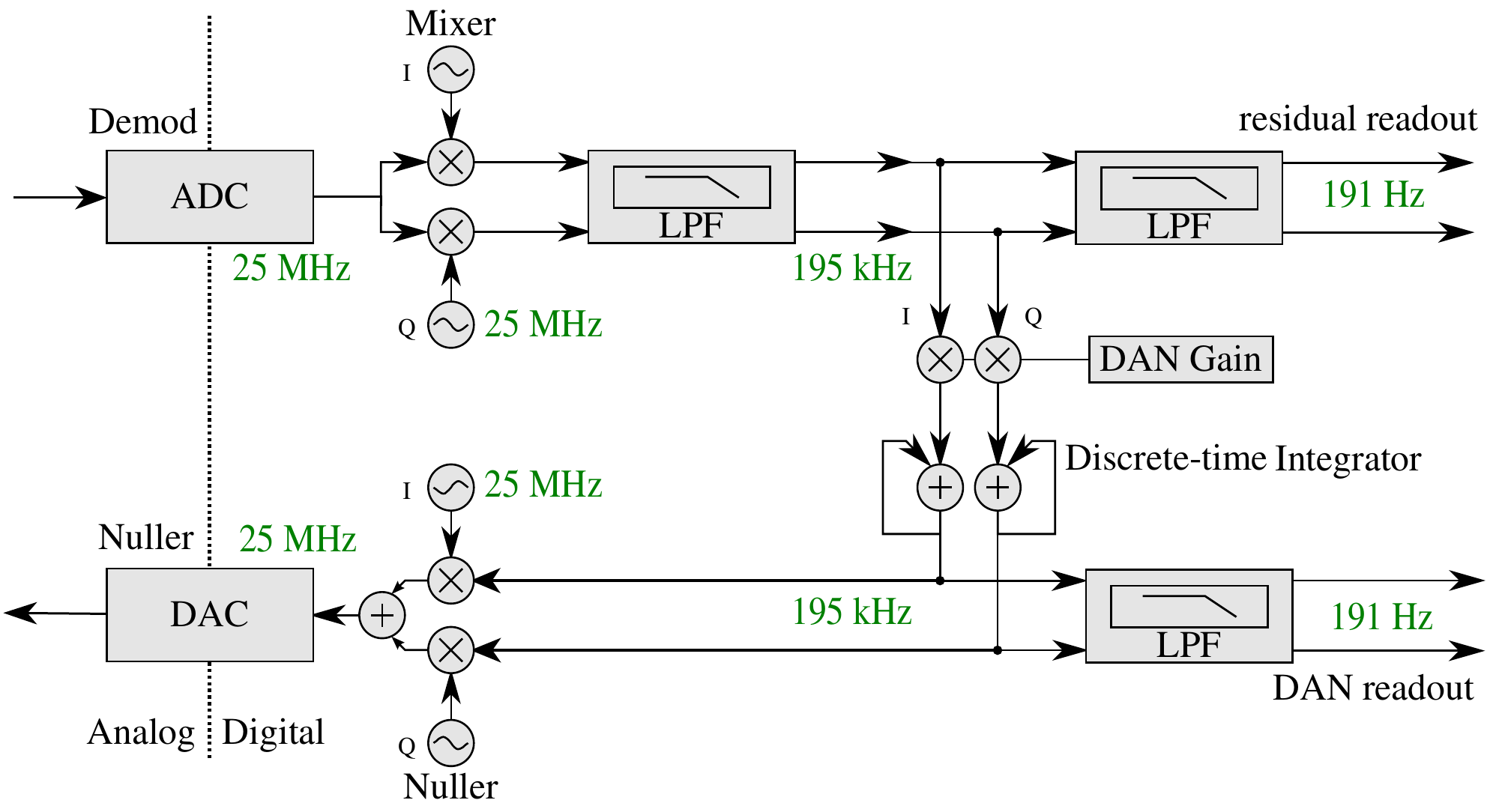}
  \caption{Schematic diagram showing the firmware portion of the DAN signal path. Data rates are shown in green. Both readout paths are shown. The residual readout path can be used to characterize the properties of the loop, though as long as the loop is operating correctly, the DAN readout alone respresents the bolometer current.}
  \label{fig:dan_path}
\end{figure}

DAN is implemented as a separate discrete time integral control loop at each bias frequency. For one such bias frequency, the feedback loop is shown in Figure \ref{fig:dan_path}. We choose integral control for its desirable property of having an effective loop gain proportional to $1 / \delta f$, where $\delta f$ denotes the frequency separation from the bias frequency. This implies that it has infinite gain at $\delta f = 0$, where the bulk of the signal (the bias voltage) resides. For stability, we require that the open loop gain falls below unity at 180\degree\ of phase shift, such that no poles of the closed loop transfer function have a positive real component. The phase shift is dominated by digital time delays, which, in our implementation, range from $5.5~\mathrm{\mu s}$ - $10.6~\mathrm{\mu s}$ depending on the bias channel. Thus, we can easily achieve stability by letting the open loop gain fall below unity at $\sim 10~\mathrm{kHz}$, which implies that at the highest bandwidth that may be required 
for science, say $100~\mathrm{Hz}$, we still have a loop gain of 100 and hence $99\%$ effective nulling.

\subsection{Experimental Results}

\begin{figure}
  \begin{minipage}[b]{0.45\linewidth}
  \centering
  \includegraphics[width=\linewidth]{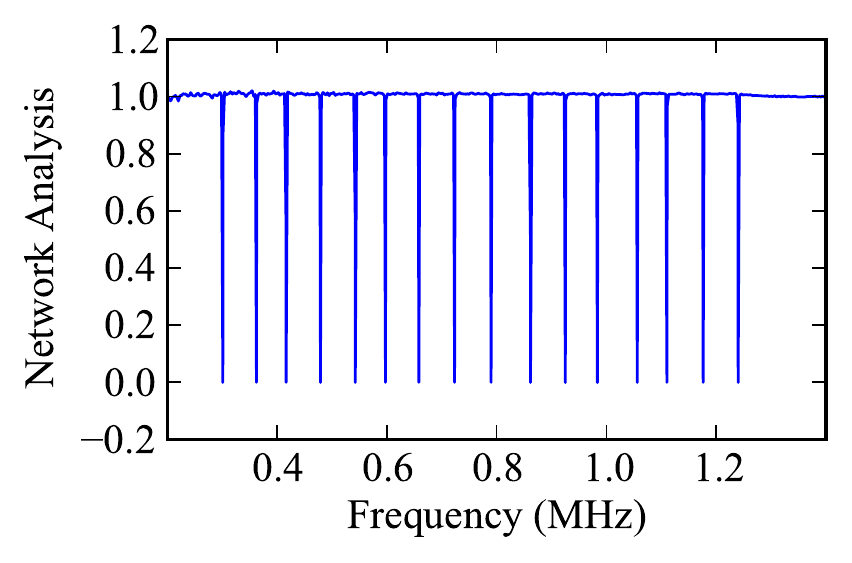}
  \end{minipage}
  \hspace{1.0cm}
  \begin{minipage}[b]{0.45\linewidth}
  \centering
  \includegraphics[width=\linewidth]{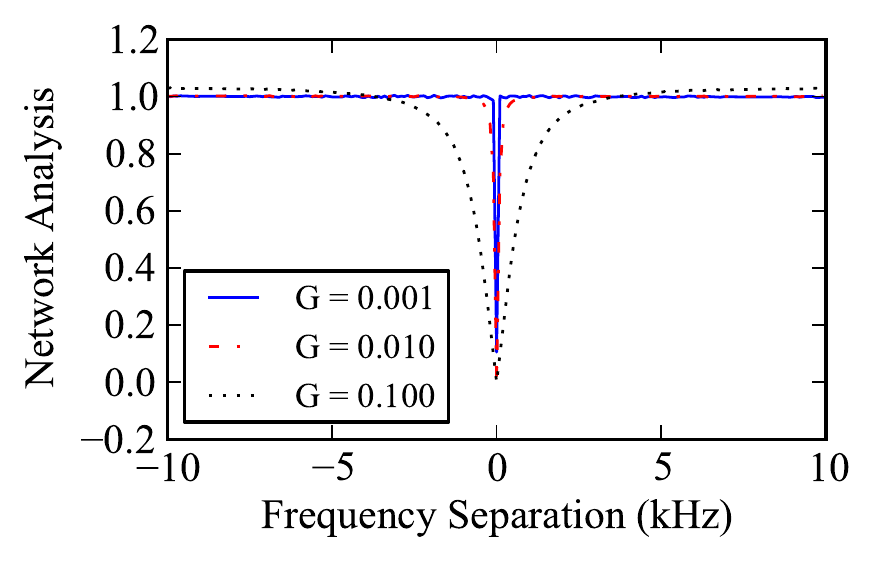}
  \end{minipage}
  \caption{Left: Network analysis showing the broadband closed loop response of DAN. At 16 arbitrarily chosen bias frequencies where DAN is operating, the transfer function of an injected signal dips to zero i.e. the signal is nulled. Right: Zoom on the network analysis near one bias frequency for a variety of gain settings. The effective bandwidth of the nulling loop is shown to increase with G, a commanded gain parameter proportional to the open loop gain of the DAN feedback loop.}
  \label{fig:dan_netanal}
\end{figure}

We initially characterized our implementation of DAN by bypassing the cryogenic components of the readout system and using a warm resistive network that has a similar transfer function to that of the cryogenic components. A measurement of the residual signal at the current sensor as a function of frequency (hereafter referred to as a network analysis) of DAN in this characterization setup reveals the closed loop transfer function shown in Figure \ref{fig:dan_netanal}.

We then operated DAN with cryogenic bolometers, including the SQUID readout. We found that one advantage of the DAN mode of operation is that the time required to map out the TES transition is reduced to the trivial exercise of stepping in bias amplitude while recording the DAN output. This mapping is performed after each cryogenic cycle on the SPTpol experiment \cite{mcmahon09}. Using DAN decreases this tuning time (which previously took tens of minutes) by an order of magnitude, while simultaneously measuring the voltage-resistance characteristic at a much higher resolution. This speedup, simplification and improvement of the bolometer tuning algorithm has been implemented and is currently one of the uses of DAN occurring daily on SPTpol, and is planned to be used on the EBEX balloon-borne experiment \cite{reichborn10}.

\begin{figure}
  \centering
  \includegraphics[width=0.49\linewidth]{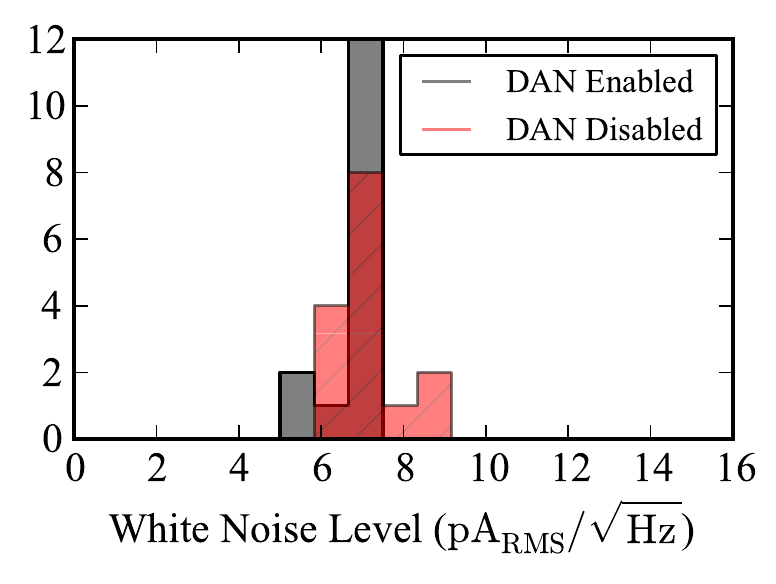}
  \caption{Histogram comparing the white noise level for 15 DAN channels operating on a SQUID, spaced between 200~kHz and 1.2~MHz, with DAN enabled and DAN disabled. The two distributions are statistically consistent, confirming our expectation that the DAN operation does not alter the noise properties of the readout system appreciably.}
  \label{fig:noise_comparison}
\end{figure}

In the cryogenic test setup, we verified that the loop is operating correctly by streaming the residual demodulator signal and seeing a highly suppressed white noise level that increases with frequency in the manner expected from the integral control loop. We also verified that the DAN signal is the same as the demodulator signal when the gain parameter is set to zero. In particular, we provide a small sinusoidal test signal, which applies time varying power to the TES. The resulting demodulated current measurement\cite{lueker09} is found to be identical between the cases where DAN is enabled and where it is disabled. In addition, we compare the white noise level between these two modes of operation and find that they agree. A histogram of the white noise levels for a cryogenic SQUID without TES bolometers attached is shown in Figure \ref{fig:noise_comparison}. 

\subsection{Improvements due to DAN}

Operating the system in DAN mode has several advantages. First, the strongly suppressed SQUID input impedance results in stronger voltage bias, improving TES stability. 

DAN nulls the SQUID current near bias frequencies, improving SQUID linearity and stability due to the large reduction of the total signal amplitude at the SQUID input. Similarly, improved nulling reduces the dynamic range requirement on the ADC, avoiding overloads in the case of unstable bolometers or changes in the bolometer operating point.

Removal of the \fll improves system stability, allowing longer wire lengths between the 4K SQUID and room temperature electronics, and increasing the allowed bandwidth. Additionally, the \fll instability to a single superconducting leg in the comb is eliminated. DAN does not interact across bias frequencies, so any instabilities are localized to one TES, instead of the whole comb. We have experimentally verified this by allowing various numbers of individual bolometers to go superconducting and found no measurable effect on the noise level of the remaining bolometers on the same comb.

The increase in usable bandwidth, removal of sensitivity to single misbehaving bolometers, and reduced usage of SQUID dynamic range, all enable a higher multiplexing factor. This higher multiplexing factor and the longer allowed wiring length are key for reducing heat load on cold stages such that it falls within the capabilities of available space cryogenics, as well as enabling lower cost, higher pixel count ground based experiments.

Both increasing wiring lengths and moving the bolometer channels to higher frequency causes an increase in the wiring inductance, which is a source of series impedance to the TES. We note that the presence of this inductance shifts the optimal bias frequency and that at this frequency, the resulting effective impedance in series with the TES is smaller than $j \omega L_\mathrm{wiring}$. The amplitude and phase of the effective complex series impedance are non-trivially related to the details of the LCR network.

The series impedance due to wiring can be mitigated by moving this wiring inside the DAN feedback loop, suppressing it in a manner similar to the SQUID input impedance suppression. This involves increasing the number of wires from each SQUID to a cryogenic stage near the detectors from two to six. An alternate scheme to mitigate this wiring inductance is DEVB, which we discuss in \S\ref{sec:devb}.

\section{Digitally Enhanced Voltage Bias}
\label{sec:devb}

While the SQUID input impedance is strongly suppressed by DAN, residual series impedance (e.g. from wiring inductance, capacitor ESR, or magnetic coupling to lossy material) can limit TES stability. This series complex impedance can be measured to good accuracy by tuning the bolometers into their transition, cooling them below their superconducting temperature, and stepping down the bias voltage until the TES latches into its superconducting state. At this point, the current is proportional to the applied voltage (as we have experimentally verified),  giving a measure of the residual complex impedance. We will show how this impedance can be used to dynamically adjust the voltage bias such as to keep the TES voltage constant.

TES bolometers respond to incident power with a time constant $\tau_0 = C/G$ where $C$ is the heat capacity of the TES and $G$ is the thermal conductance to the heat sink. Under voltage bias, this time constant is sped up by the loop gain of electrothermal feedback \loopgain \cite{richards94} as
\begin{equation}
  \label{eq:tau}
  \tau_\mathrm{eff} = \frac{\tau_0}{1+\loopgain} \, .
\end{equation}
DEVB actively controls the voltage across the TES on time scales faster than $\tau_\mathrm{eff}$, providing a constant voltage bias across the TES while strongly suppressing the effective series complex impedance.

We denote this series impedance as $z_s = R_s + j X_s$. Given values of $R_s$ and $X_s$, a constant voltage bias to the TES denoted $V_\mathrm{TES}$ can be provided by measuring the current through the bolometer and providing the voltage
\begin{equation}
  \label{eq:daf}
  V_\mathrm{OUT} = \sqrt{(V_\mathrm{TES} + R_s I)^2 + X_s^2 I^2} \, ,
\end{equation}
where $I$ denotes the magnitude of the measured current and $V_\mathrm{TES}$ is the programmable voltage set-point. Since, as with DAN, there are digital delays between measuring $I$ and providing $V_\mathrm{OUT}$, we perform this computation at a fixed loop gain below unity in order to introduce an effective DEVB time constant and ensure stability of the feedback loop. The DEVB time constant is discussed further in \S\ref{sec:implementation}.

In addition to providing improved TES stability, the enhanced voltage bias across the LCR comb suppresses cross-talk due to that same series impedance. In the case of an inductive series impedance, this term is proportional to the ratio of the stray inductance to the inductance of the LC resonator\cite{dobbs11}. With DEVB, this effect is suppressed by the DEVB loop gain.

\subsection{Preliminary Implementation}
\label{sec:implementation}

A hardware prototype of DEVB has been implemented based on a customized version of the DfMux firmware\cite{smecher10}. This firmware supports 64 bias and nuller channels, as well as 68 single phase demodulator channels. The DEVB feedback loop is implemented on one of those bias channels, using two of the demodulator channels, aligned 90\degree out-of-phase to provide I and Q current measurements at the bias frequency. The resulting complex signal is fed to a DEVB module, which calculates an output amplitude for the synthesizer block by explicitly computing Equation \ref{eq:daf}, using the quadrature sum of the I and Q measurements as the current amplitude. The synthesizer modulates the amplitude of the bias sinusoid using this output. The input to, and the output from, the DEVB module are further decimated to 191 Hz and streamed across the network for setup and analysis.

\begin{figure}
  \centering
  \includegraphics[width=\linewidth]{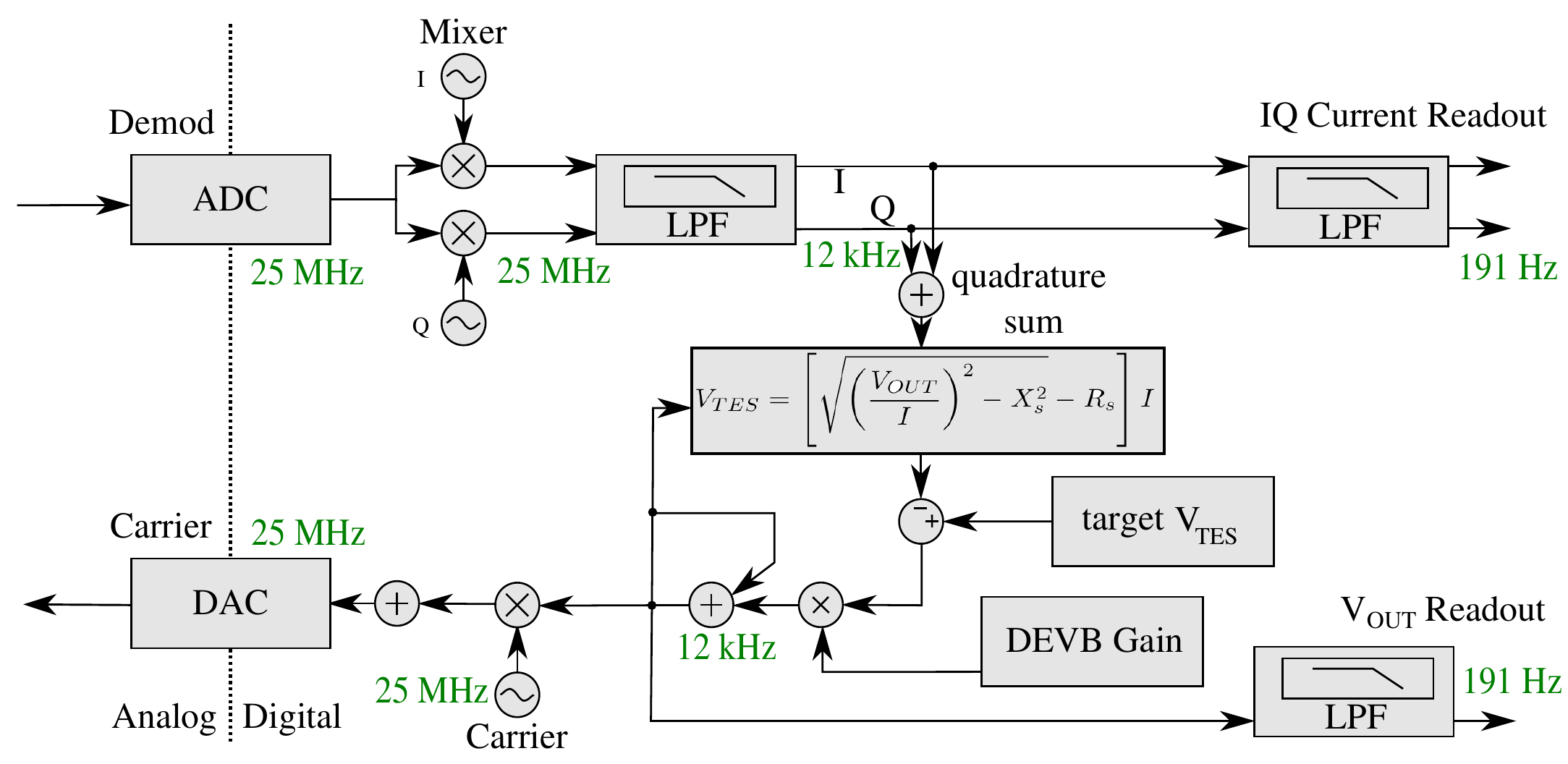}
  \caption{Firmware diagram showing the DEVB signal path. Data rates are shown in green. The complex current measurement and output voltage amplitude are read out simultaneously.}
  \label{fig:daf_path}
\end{figure}

Figure \ref{fig:daf_path} shows the computation within the DEVB module. At the input to this module, fixed-point signals from the demodulator are converted into floating-point numbers for computation. The algorithm itself is separated into two parts: calculation of a target amplitude, and a control element generating the feedback signal by stepping the current amplitude a fraction towards the target amplitude. We denote this fraction the DEVB gain. The control signal generated by these blocks is finally converted back into fixed-point numbers suitable for the synthesizer block.

We run the control loop at 12 kHz and set the DEVB gain to $1/3$, yielding an approximate DEVB time constant of 4 kHz; much faster than the $\sim 100~\mathrm{Hz}$ time constant of a typical TES.

\subsection{Measurements}
\label{sec:measurements}

\begin{figure}
  \hspace{0.5cm}
  \begin{minipage}[b]{0.4\linewidth}
  \centering
  \includegraphics[width=\linewidth]{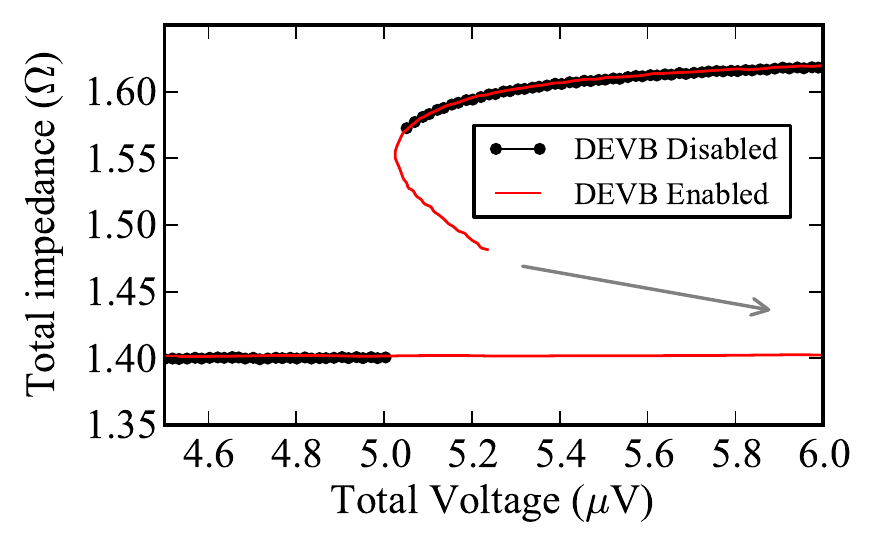}
  \end{minipage}
  \hspace{1.5cm}
  \begin{minipage}[b]{0.4\linewidth}
  \centering
  \includegraphics[width=\linewidth]{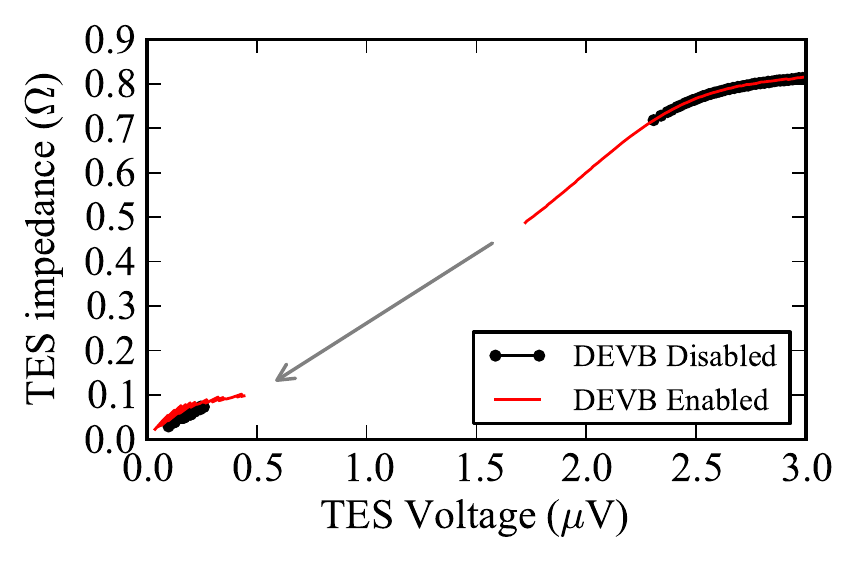}
  \end{minipage}
  \caption{TES transition shown as the provided voltage is slowly decreased with DEVB enabled and disabled. The TES latches into a superconducting state when it reaches $60\%$ of its normal resistance with DEVB enabled, compared to latching at $90\%$ of its normal resistance with DEVB disabled. The left panel shows the raw data, whereas the right panel shows the data once the stray series impedance is subtracted off. The arrows are used to show the latching event, which occurs faster than the data rate.}
  \label{fig:vr}
\end{figure}

In order to demonstrate the enhanced performance of TES detectors under DEVB, we test the stability of a detector in the presence of a large stray impedance as given in Equation~\ref{eq:stability}. Specifically, we apply a large carrier voltage to a single detector in an LC resonance, designed to allow for frequency multiplexing. By setting the frequency of this voltage significantly away from the LC resonance we create an impedance in series with the bolometer. In this case we set the series impedance to $1.4~j~\Omega$, while the normal TES resistance is $\sim0.9~\Omega$. We then reduce the applied voltage, dropping the TES into its transition. The TES is lowered through its transition until it latches and becomes superconducting. The resulting total impedance as a function of applied voltage and the resistance of the TES alone as a function of the voltage across the TES alone are shown in Figure \ref{fig:vr}. This test, performed with DEVB enabled and DEVB disabled, shows that the use of DEVB lowers the 
point at which the TES latches, demonstrating that DEVB allows the TES to drop much lower into its transition in the presence of large series impedance.

\subsection{Future Prospects for DEVB}

While we have demonstrated the concept of DEVB, extending this feedback mechanism to an array of bolometers requires several improvements to the firmware, as well as the readout scheme and feedback mechanism.

Equation \ref{eq:daf}, as well as the computation of the magnitude of the complex demodulated current, contains a square root operation, which is challenging to implement efficiently on an FPGA. An alternate DEVB scheme would involve controlling both the I and Q components of the bias voltage $V_\mathrm{OUT}$ rather than controlling only its amplitude. The set point for $V_\mathrm{OUT}$ would then become
\begin{equation}
 \begin{split}
  V_\mathrm{OUT}^I = V_\mathrm{TES}^I + I^I R_s - I^Q X_s \\
  V_\mathrm{OUT}^Q = V_\mathrm{TES}^Q + I^Q R_s + I^I X_s \, .
 \end{split} 
\end{equation}

This complex DEVB scheme, when implemented in fixed-point logic, would be easily scalable to the full 64 bias and complex demodulation channels using the current Xilinx Virtex-4 FPGA. 

The current implementation of DEVB does not support DAN. In future versions of DEVB, we would implement both the DEVB and DAN modules simultaneously. We would choose to run DAN at a significantly faster effective time constant than the DEVB control loop, such that the SQUID current is nulled, regardless of the changes in $V_{\mathrm{OUT}}$ made by the DEVB module. This approach would allow us to employ the benefits of DAN, such as the higher usable bandwidth and reduced SQUID dynamic range requirement, while retaining the suppressed series impedance from wiring inductance without needing to increase the heat load from the 4K stage to the millikelvin stage by increasing the wire count.

Finally, it is possible to implement DEVB with frequency-dependent coefficients such that a frequency-dependent series impedance could be suppressed. This would allow narrower LC filters to be used, increasing the multiplexing factor for a given usable bandwidth and constraint on electrical cross-talk.

\section{Conclusion}

We have introduced DAN, a method of keeping the current through the sensing portion of the readout system very close to zero at frequencies near the TES bias frequencies. This digital feedback mechanism increases stability and linearity of the SQUIDs, while suppressing their effective input impedance. We have demonstrated a fully implemented version of DAN including stability results and nominal noise performance, and described its use on the SPTpol experiment.

While this technique is similar to BBFB, the suppression of series impedance is unique to DAN. Removing the \fll from the readout system increases the usable bandwidth for frequency multiplexing, enabling higher multiplexing factors and reducing wiring length constraints. Hence, DAN is a key technology for applying frequency multiplexed readout to satellite applications.

However, the use of higher bias frequencies increases the series impedance due to wiring. This effect can be mitigated by increasing the number of wires per multiplexing module from the 4 K SQUID to (close to) the millikelvin stage from two to six. 

We also described DEVB, a digital feedback technique which suppresses series impedances to the TES by measuring them and providing the actively controlled bias voltage required to hold the TES at a constant potential. This technique has been demonstrated to increase TES stability, despite large ($z_s \gtrsim R_{\mathrm{TES}}$) complex series impedances, without increasing wire count.

\appendix

\section{Stability criterion for series impedance to the TES}
\label{app:stability}

The left hand panel of Figure \ref{fig:vr} shows that the reason for the TES latching due to series impedance is well understood. The total impedance is seen to have a zero derivative with respect to the total voltage, so further lowering the total bias voltage causes the voltage across the TES to fall in a runaway manner. This instability occurs when 
\begin{equation}
  \label{eq:instability_crit}
  \frac{d | z_s + R_{\mathrm{TES}} |}{d V_{\mathrm{OUT}} } = 0 \, .
\end{equation}

We substitute Equation \ref{eq:daf} which describes the model, and assume that $R_{\mathrm{TES}}$ is a monotonically decreasing function of $V_{\mathrm{TES}}$. We find that this condition is met when
\begin{equation}
  \label{eq:stability_app}
  \frac{d \log V_{\mathrm{TES}}}{d \log R_{\mathrm{TES}}} = \frac{|z_s|}{|z_s + R_\mathrm{TES}|}
\end{equation}
and conclude that the TES will latch due to series impedance when the fractional stray impedance exceeds the logarithmic slope of the TES transition.

\acknowledgments

We acknowledge funding from the Canadian Space Agency, Natural Sciences and Engineering Research Council, Canadian Institute for Advanced Research, and Canadian Foundation for Innovation. MD acknowledges support from an Alfred P. Sloan Research Fellowship and Canada Research Chairs program. 

We would like to thank the POLARBEAR and EBEX collaborations and in particular Kam Arnold and Ben Westbrook for their work on the TES bolometers used in this work.

\bibliography{dan_daf_spie2012.bib}
\bibliographystyle{spiebib}
\end{document}